\documentclass[conference]{IEEEtran}
\IEEEoverridecommandlockouts

\usepackage{amsmath,amssymb,amsfonts,amsthm,mathtools}
\usepackage{physics}

\usepackage{graphicx}
\graphicspath{{.}{figures/}}
\usepackage{xcolor}

\usepackage{tikz}
\usetikzlibrary{positioning}
\tikzset{%
    block-common/.style={draw, semithick, fill=white, minimum height=2em, minimum width=2em},
    block/.style={rectangle, block-common},
    txtblocktight/.style={block, align=center, minimum height=2.3em},
}

\usepackage{qcircuit}

\usepackage{cite}
\usepackage{booktabs}
\usepackage[capitalize]{cleveref}

\newtheorem{theorem}{Theorem}

\theoremstyle{remark}
\newtheorem{remark}[theorem]{Remark}
\newtheorem{observation}[theorem]{Observation}

\let\eps\varepsilon
\let\varepsilon\epsilon
\let\epsilon\eps

\usepackage[draft,nomargin,marginclue,inline]{fixme}

\makeatletter
\renewcommand*\FXLayoutMarginClue[3]{%
  \marginpar[%
  \raggedleft\@fxuseface{margin}\textcolor{black}{\ignorespaces#3 fixme}]{%
    \raggedright\@fxuseface{margin}\textcolor{black}{\ignorespaces#3 fixme}}}
\makeatother

\begin{document}

\title{Implementation and Experimental Evaluation of Reed-Solomon Identification}

\author{%
    \IEEEauthorblockN{%
    Roberto Ferrara\IEEEauthorrefmark{1}, 
    Luis Torres-Figueroa\IEEEauthorrefmark{2},  
    Holger Boche\IEEEauthorrefmark{2}, 
    Christian Deppe\IEEEauthorrefmark{1},\\ 
    Wafa Labidi\IEEEauthorrefmark{2}, 
    Ullrich M\"onich\IEEEauthorrefmark{2}, 
    Vlad-Costin Andrei\IEEEauthorrefmark{2}
    }
    \IEEEauthorblockA{
    \IEEEauthorrefmark{1} 
    Institute for Communications Engineering,
    \IEEEauthorrefmark{2} 
    Chair of Theoretical Information Technology\\
    Technical University of Munich, D-80333 Munich, Germany\\
    Email: \{roberto.ferrara, luis.torres.figueroa, boche, christian.deppe, wafa.labidi, moenich, vlad.andrei\}@tum.de}
}

\maketitle

\begin{abstract}
Identification is a communication paradigm that promises exponential advantages over transmission for applications that do not actually require all messages to be reliably transmitted.
Notably, the identification capacity theorems prove exponentially larger rates compared to classical transmission.
However, there exist additional trade-offs that are not captured by these theorems and which become relevant for the deployment of identification in practical communication settings. In particular, in this paper we evaluate the latency introduced by computations at the encoder and decoder when employing identification codes. 
For this, we implement them using an explicit code construction based on Reed-Solomon codes and integrate it into a single carrier transmission system using software-defined radios. Our evaluation of the practical aspects of identification codes  show that unless care is taken, these trade-offs can compromise the theoretical advantage given by the exponentially large identification rates.
\end{abstract}


\section{Introduction}
    Shannon's transmission theory of communication~\cite{shannon} along with technological advancements have allowed unparalleled progress in communication systems and increased the speed of information transmission.
    Much of this progress is still ongoing today~\cite{5GShannon} and only recently  it has started to reach its own limits in some areas~\cite{6gnetworks}.
    While Shannon's transmission models have been traditionally applied in any setting where messages need to be reliably transmitted from a sender to a receiver, there do exist other models of communication beyond this.
    Such models are collectively called post-Shannon theory \cite{CBDSSF21}, 
    and identification is one such example \cite{AD89,AD89feedback}.
    
    In identification, the goal is to safeguard the receiver's ability to reliably verify whether he has chosen 
    the same identity as the one used by the sender for the communication, while renouncing its ability to decode the content of it.
    If the receiver were additionally tasked with reliably retrieving the \textit{value} of such identity, then we would fall back into the classical transmission scenario discussed before. 
    Thus, we can think of identification as a relaxation of transmission. 
    Namely, transmission codes satisfy the conditions of identification codes, but not vice versa.
    
    One of the main advantages of the identification paradigm lies in its scalability.
    That is, as we move onto identification codes that do not allow for transmission, or otherwise said, if we sacrifice  the receiver's ability to decode and only preserve its ability to verify, then we find an exponential increase in the number of identities that can be reliably verified. More precisely, this means a doubly-exponential number of identities in the block length of the channel~\cite{AD89}.
    
    When regarded in the opposite direction, i.e., for a given set of identities and messages of the same size, this result translates into an exponential reduction in the needed block length to be sent when using identification compared to the transmission case.
    Thus identification can potentially find application scenarios within control systems~\cite{controlbook}, watermarking~\cite{MK06, AHLSWEDEwater}, the automotive domain~\cite{autoBoche}, recommendation systems~\cite{DDF20}, and in general in any setting characterized by the need for quick or small checks, leading to unprecedented capabilities and the reduction of bottlenecks, as well as channel congestion.
    
    Despite these promising theoretical results, the practicability of identification still needs to be demonstrated. This paper thus represents an additional contribution towards this objective. Here, we take the composition of Reed-Solomon codes introduced in \cite{VW93explicit}, as described in \cite{DDF20}, and use it to perform identification across wireless communication. 
    Except for a subset of parameters with fast computation time,
    we find that the computational overhead in terms of latency introduced by this approach is a few orders of magnitude above the transmission time.

    Although the identification code construction that we analyze herein achieves the aforementioned double exponential increase, it does so at a high computational cost, as shown by our results. This stresses the notion that theoretical code constructions that fulfill the requirements posed by identification must be additionally accompanied with performance analysis of their implementations in order to assess their feasibility for integration into practical applications. Further, our results also emphasizes the need to find other code constructions that can be  efficiently implemented, such as the work presented in \cite{SFD21}.
    
    The main contribution of this paper are therefore twofold:
    \begin{enumerate}
        \item We integrate a Reed-Solomon implementation of identification codes into an experimental wireless setup and evaluate its suitability for practical applications
        \item We assess the system-level performance overhead introduced by such integration and identify practical aspects that are relevant for future implementations
    \end{enumerate}

    The paper is structured as follows. In \cref{sec:identification} we overview identification, its coding, and the identification codes considered here.
    In \cref{sec:multiple}, we analyze the scenario of multiple parties concurrently using identification.
    Our hardware setup is described in \cref{sec:experiment}, while in \cref{sec:single} we analyze the results of our experiments.
    Final conclusions and further discussion are finally presented in \cref{sec:conclusion}.

\section{Identification}
\label{sec:identification}

\subsection{System Model}

    Identification can be reduced to transmission via a pre-processing step at the sender and a post-processing step at the receiver~\cite{AD89}, which without loss of generality we can think of as a random ``challenge'' sent by the sender with identity $i$ and verified at the receiver with identity $j$.
    In order to create the challenge, to each identity $i$ is associated a function $t_i$.

\begin{figure}
      \centering
      \includegraphics[width=\linewidth]{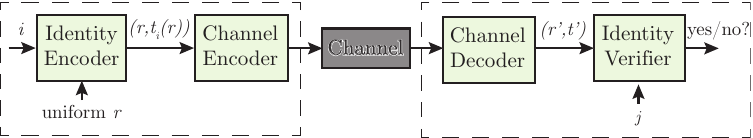}
      \caption{System model for identification.}
      \label{fig:system_model_id}
     \end{figure}

    The challenge is composed of a randomly chosen input $r$ (the randomness) and the output $t_i(r)$~\cite{AD89feedback}, which we call a tag~\cite{DDF20}. This randomness-tag pair is the challenge and is sent to the receiver via a transmission code (to avoid transmission errors) which outputs $r',t'$. The receiver with identity $j$ then verifies the challenge by recomputing the tag on the received randomness $r'$ with $t_j$, who concludes that $j=i$ (accept) if the computed and received tags are the same ($t_j(r')=t'$), and concludes that $i\neq j$ (reject) otherwise. The system model used here that summarizes these relationships is shown in \cref{fig:system_model_id}.
    
    The collection of these functions is sometimes itself known as an identification code, since it alone defines the whole identification code if the channel is noiseless. We will thus refer to it as a noiseless identification code, or $0$-ID code since it corresponds to $\eps_1=0$ below.

\subsection{Mathematical Notation}

    Formally, such an identification code looks as follow.
    We use the notation $[N]\coloneqq\{0,...,N-1\}$ for any integer $N>0$ and $WV(z|x) = \sum_y W(y|x) V(z|y)$ for channel composition.
    Let $I$, $R$ and $T$ be integers and let $M = RT$. 
    Let $\{t_i:[R]\to[T]\}_{i\in[I]}$ be a collection of functions such that for any two functions the fraction of overlaps (collisions) is bounded by $\epsilon_2\leq 1$, namely such that for any $i\neq j\in [I]$ we have $\sum_{r\in [R]} \delta_{t_i(r), t_j(r)}/R \leq \epsilon_2$, where $\delta$ is the Kronecker delta.
    This defines an $(n=\log M,I,\epsilon_2)$
    $0$-ID code via the $n$-bits noiseless identity channel, where for every identity $i\in[I]$ the stochastic encoder $E(m|i)$ picks $r\in[R]$ uniformly at random and sends the challenge $m = (r, t_i(r))$ and the verifier $V(v|(r,t),j)$ accepts ($v=1$)  if $t_j(r)=t$ and rejects ($v=0$) otherwise.
    The false reject error (concluding that the identities differ even though they are equal) and false accept error (concluding that the identities are the same even though they are different) satisfy
    \begin{align*}
    EV(0|i,i) &=0 \leq \epsilon_2 ,
    &
    EV(1|i,j) &\leq \epsilon_2,
    & 
    \forall\ & i \neq j.
    \end{align*}

    In the general case of a noisy channel, a transmission code with error $\epsilon_1$ can be concatenated with the $0$-ID code to obtain an $(n,I,\epsilon_1 + \epsilon_2)$ identification code.
    The most important result in identification theory is that such ID codes allow to achieve an identification rate equal to the transmission rate and thus the identification capacity is as large as the Shannon capacity of the channel.
    However, the rate of an identification code is defined as $\frac{1}{n} \log \log I$ instead of $\frac{1}{n} \log I$, thus the identification capacity theorem says that the number of identities that can be achieved while sending the errors to zero is doubly exponential in the block length with rate equal to the Shannon capacity. 

\subsection{Explicit Constructions of Identification Codes}

    Given a capacity achieving transmission code, explicit constructions of capacity achieving identification codes are known~\cite{VW93explicit}.
    The double exponential growth means that already at practical block lengths the identification code sizes will be beyond the size of the universe.
    But it also means that for the encoder and verifier to access an identity, they need to access an exponential number of bits, which is already computational inefficient without counting any computation after that.
    Still, the result can also be used in the opposite direction: for a given code size, we see an exponential reduction in block length with respect to transmission.
    The goal is to show this advantage practically.
    
    \newcommand{\RST}{{\text{RS2}}}
    As observed in~\cite{DDF20} the $0$-ID codes are in one-to-one correspondence to error correction codes (ECC), by mapping any codeword $c_i$ to the function $t_i(r) = c_{ir}$ outputting the symbols of the codeword when given a symbol location, and vice versa.
    A minimum distance $d$ on the ECC is then a bound $\epsilon_2 = 1- d/M$ on the fraction of collisions of the $0$-ID code.
    For Reed-Solomon codes, that are already themselves generated by the output of polynomials over finite fields, the identity functions are these polynomials~\cite{VW93explicit,MK06, DDF20}.

    In this paper we continue the study of the $0$-ID code introduced in \cite{VW93explicit}, obtained by concatenating two maximal sizes Reed-Solomon codes, moving from prime fields to bitstring fields.
    Let $[q,k]_{RS}$ be the Reed-Solomon code over the field of size $q$ generated by the polynomials with degree less than $k$. 
    Given parameters $q>k>\delta$, we concatenate, as explained in \cref{remark:concatenation} below, a $[q^k,q^{k-\delta}]_{RS}$ with a $[q,k]_{RS}$ to obtain $I = q^{k q^{k-\delta}}$ functions from $\mathbb{F}_{q^k} \times \mathbb{F}_q \to \mathbb{F}_q$ (thus $R=q^{k+1}$ and $T=q$), with $1-\epsilon_2 = (1 - \frac{k+1}{q}) (1 - \frac{1}{q^{\delta}} + \frac{1}{q^{k}}) $.
    We call it the RS2 $0$-ID code.
    A challenge thus has $k+2$ symbols in $\mathbb{F}_q$, with $k+1$ of randomness and only one of tag. 
    
    \begin{remark}
    \label{remark:concatenation}
    \newcommand{\inner}{{\prime\prime}}
    \renewcommand{\outer}{{\prime}}
    We can view the concatenation of ECCs
    as a concatenation of $0$-ID codes. 
    Namely, if $c_i^\outer$ is a codeword of blocklength $R^\outer$ and every symbol of $c_i^\outer$ is encoded by the codewords $c_j^\inner$ of blocklength $R^\inner$, then the resulting codeword and function $c_i$ from the concatenation are
    \begin{align}
        c_i 
        &= c^\inner_{ c^\outer_{i1}} \cdots c^\inner_{c^\outer_{iR^\outer}},
        &
        c_i(r^\outer, r^\inner) &= c^\inner_{ c^\outer_{ir^\outer},r^\inner},
    \end{align}
    where the codeword has block length $R^\outer \cdot R^\inner$ and the randomness pair $r^\outer, r^\inner \in [R^\outer]\times [R^\inner]$ is the input to the function.
    Functional composition is recovered if we let the randomness itself be a function on the codewords as $r(c_i)\coloneqq c_{ir}$, then:
    \begin{equation}
        c_i(r^\outer, r^\inner) = r^\inner (r^\outer(c_i)) = r^\inner \circ r^\outer(c_i).
    \end{equation}
    The challenges of such concatenated $0$-ID codes then look like
    \begin{equation}
        \left(r^\outer, r^\inner, r^\inner \circ r^\outer(c_i)^{\vphantom{A}}\right),
    \end{equation}
    namely a sequence of random inputs, one for each code in the concatenation, with a final concatenation-computed tag. 
    Since the size of identities can be a drawback, such a trick can allow the tag to be computed as the identity is being streamed in the verifier, as is the case for Reed-Solomon $0$-ID codes.
    \end{remark}

\section{Scenarios involving multiple receivers}
\label{sec:multiple}
\newcommand{\nchallenges}{{n_\text{c}}}
    We can use the identities to encode data or identify multiple receivers. A positive verification can then be used to trigger further action.
    For example, we could use an identification code to hail a receiver among many, and subsequently the stream of status data from the identified receiver.
    However, not all such scenarios display an advantage over simply transmitting a unique string identifying the receiver.
    If all possible identities verify the challenge, then exponentially many of them will verify a false positive.
    Since the tag is the only element differentiating the identities in the verification process, the average fraction of identities accepting a tag is exactly $1/T$.
    Indeed, let $I_t/I$ be the fraction of identities producing the tag $t$ at a certain randomness, then the average of the possible tags is exactly $\frac{1}{T}\sum_t I_t/I = 1/T$.

    For the identification codes that we consider, the situation is even simpler: the fraction of identities that accept any given tag is exactly $1/T=1/q$.
    This means that for any number $\nchallenges$ of challenges sent, the fraction of identities that will accept will be exactly $1/q^\nchallenges$, with only one of them being a true accept.
    This is a general property of identification codes created by maximum-distance separable codes, as shown in the following observation.
    \begin{observation}
    To show that the fraction of identities is always uniformly $1/q^\nchallenges$, consider any full rank $\nchallenges \times k$ matrix with $\nchallenges \leq k$.
    The range of such matrix is $(\mathbb{F}_q)^\nchallenges$ and every element in it appears exactly $q^{k-\nchallenges}$ times, which is the size of the the null space of the matrix, because every input $x$ can be written as $\hat{x} + x_0$ with $\hat{x}$ in one-to-one with $(\mathbb{F}_q)^\nchallenges$ and $x_0$ any element sent to zero.
    Since sending a set of tags means sending the output of a submatrix of the generator matrix of the Reed-Solomon code, and since for it (or more generally for any maximum-distance separable code) any $\nchallenges\leq k$ columns of the generator matrix are linearly independent~\cite[Theorem 5.3.7, p. 237]{Roman92coding}, we obtain that any set of tags appears in the same number of identities.
    \end{observation}
    For the RS2 $0$-ID code used here each challenge is $k+2$ symbols in $\mathbb{F}_q$, while a unique identity is determined by $k \cdot q^{k-\delta}$ symbols.
    This means that
    \begin{equation}
        \frac{k}{k+2} q^{k-\delta}
    \end{equation}
    challenges can be generated before it becomes shorter to simply send the $k \cdot q^{k-\delta}$ symbols determining the identity, but the penalty incurred by computing the multiple identities will degrade any advantage long before reaching this bound.

    If the goal is to avoid any false positive and the set of receivers is fixed, then identification cannot be an advantage over transmission in this multiple-receivers scenario.
    Indeed, for an advantage to exist the number of receivers must be at least $M=RT$ otherwise we can simply assign a unique identifier for each receiver with the same communication overhead. However, at that point the probability of false positive will be  large (especially for codes achieving identification capacity where $R\gg T$ and even $\log R \gg \log T$).
    Thus implementations such as~\cite{BringerPatent} may not show an advantage without additional considerations.
    
    Instead, identification might show an advantage if the set of receivers is a variable subset in a larger set of total possible receivers. 
    Even the size of the set of receivers itself can be variable, since we can send multiple challenges to accommodate for a larger number of receivers. A number of challenges $\nchallenges$ will see an average fraction $1/T^\nchallenges$ of identities accepting all the challenges.

\subsection{Use Case 1: Infrastructure-based V2X Communications} \label{uc1}

    A scenario displaying multiple receivers in variable numbers is the case of autonomous driving.
    Given the response time of an autonomous car and the amount of traffic, the number and the set of other cars that are relevant for consideration is variable.
    A fast identification code could allow for faster response times of the cars and prevent congestion and latency in future networks with many autonomous cars interacting with each other.
    The overhead incurred by the transmission of the randomness can be mitigated using time- or location based common randomness or pseudo-randomness.
    Much like one-time-password soft tokens, the randomness can be synchronized among all the senders and receivers, since no information about the identity is contained in the randomness.
    If all identities generate challenges on the same randomness, then by the birthday paradox this will increase the overall amount of false positives, thus a compromise between synchronized and transmitted randomness might need to be considered. 
    The use of common randomness will be subject of future work.

\subsection{Use Case 2: Next Generation Cellular Networks} \label{uc2}

    Another scenario that could benefit from identification is the next generation 6G mobile cellular network, where the number of connected devices will increase manifold in comparison with 5G networks. By using identification, not only spectral efficiency but also security in the radio access network  can be further improved.
    Under this paradigm, each user equipment (UE), regardless of whether it is a human or a machine, will be assigned an identity $i$, while its associated tag $t_i(r)$ can be used as a secure means for wireless communication with the base station (BS), similar to the classical identifiers assigned to handsets today, e.g., the International Mobile Equipment Identity (IMEI) and Subscriber Permanent and Concealed Identity (SUPI/SUCI).
    In paging scenarios, the search for a UE within the subset of identities in a tracking area code (TAC) will be void of potential collisions caused by $\eps_2$ if the NR Cell Identity (NCI) of a random BS in the TAC is used for generating the randomness $r$. Furthermore, the changing nature of $t_i(r)$ according to random NCI and TAC addresses the long-standing IMSI catching security breach in a more efficient way without employing computation-intensive encryption methods, such as the elliptic curve integrated encryption scheme in 5G. 

\subsection{Use Case 3: Deep-space communication} \label{uc3}
    As can be seen in \cref{fig:timings}, today's impressive data rates are the reason why identification might not be interesting for many of today's applications.
    One can say that a practical implementation of identification could have been interesting 20 years ago, but there at least one application where small data rates and large latency will be fundamental and that is deep-space communication.
    While identification cannot help in the transmission of images and data from a satellite or a rover, it may allow for more frequent verification of the status of the device.
    A well integrated and low power identification encoder could send a tag of the system or of the decisions made by the device and use minimal power in the communication, as the challenge only consists of a few bits (see \cref{tab:experiment_list} for some examples). 
    The control center on Earth could then verify that tags are the ones expected during correct operation and only take further action when a tag is rejected, hinting at a possible problem.

    \medskip
    
    In such applications, identification can exponentially reduce the load on the transmission network. 
    However, if power consumption and latency are also key performance indicators, then the advantage of identification is dependent on whether the challenge generation incurs less latency or power consumption with respect to simply transmitting a unique identifier.
    The computation time of the RS2 $0$-ID code was considered in~\cite{DDF20}.
    In the next section, we expand the set of explored parameters and implement it on real hardware.

\section{Experimental Setup}
\label{sec:experiment}

    Our experimental setup is implemented using multiple software defined-radios (SDR) consisting of universal software radio peripherals (USRP) 2954R  connected with servers where the bit and signal processing take place.
    In order to demonstrate experimentally the advantages of identification in wireless networks, we deploy a single-carrier transmission system. 

\subsection{Single carrier transmission system}

    Under this paradigm, the identification encoder has been abstracted at the transmitter side using an application programming interface for the challenge generation, which outputs a variable-length codeword containing the computed tag, $t_i(r)$, for constant values of $k$ and $\delta$.
    
\begin{figure}[htbp]
    \centering
    \vspace{-1.2em}
    \includegraphics[width=0.7\linewidth]{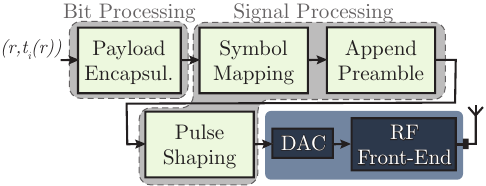}
    \vspace{-0.8em}
    \caption{Transmitter implementation using an SDR.}
    \label{fig:tx_implementation}
    \end{figure}

    This codeword is next encapsulated into a payload whose size matches the tag length at all times.
    In order to minimize regular transmission errors, i.e., errors of the first kind represented as $\eps_{1}$, a preamble is attached to each payload for countering channel effects at the receiver. Our setup uses two 13-symbol long Barker codes for frame synchronization \cite{barker53}, and pilot signals carrying a 128-symbol long Gold sequence~\cite{gold67} for channel equalization. 
     Once the frame is assembled, its preamble is modulated using BPSK, while the QPSK is used for the payload. The lower modulation order for the synchronization signals increases the robustness against errors of the first kind.
     The modulated I/Q samples are then pulse shaped at baseband before being upconverted to the carrier frequency $f_{c}$ and sent over the channel. The steps at the transmitter are shown in \cref{fig:tx_implementation}.

\begin{figure}[htbp]
      \centering
      \vspace{-1em}
      \includegraphics[width=0.96\linewidth]{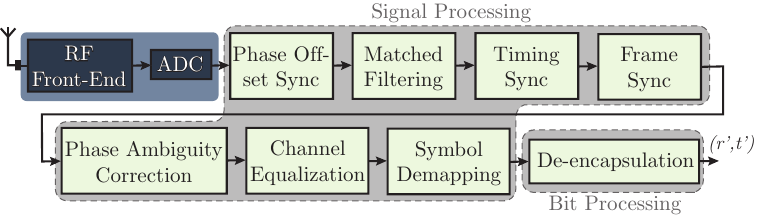}
        \vspace{-1em}
      \caption{Receiver implementation using an SDR.}
      \label{fig:rx_implementation}
     \end{figure}

     At the receiver end, the SDR filters the amplified received signal and digitalizes it at a rate $f_{s}$. 
    We use phase-locked loop based algorithms for carrier frequency and phase offset estimation before matched filtering. 
    For symbol synchronization, a timing error detector based on the zero-crossing method \cite{rice2009digital} is used, while frame synchronization is realized using a cross-correlation algorithm that exploits the autocorrelation properties of the Barker code in the preamble. 
    Once the beginning of the frame is identified, the pilot symbols are used for phase ambiguity resolution, as well as for channel estimation using a least-squares approach. 
    The equalized symbols are finally demodulated into a bitstream that is fed for the verification step in identification. Fig.~\ref{fig:rx_implementation} summarizes these procedures.

    Parallel bit- and signal-processing tasks as well as queue management for transmissions via the SDR care for a continuous operation of our setup. The parameters used in our experiments are summarized in table~\ref{tab:hw-parameters}.

    \vspace{-0.5em}
    
    \begin{table}[htbp]
    \caption{Key Parameters of the Experimental Setup}
    \centering
    \begin{tabular}{@{}lll@{}l}
      \toprule
      Parameter & Variable & Value \\
      \midrule
      Modulation schemes & $M$ & BPSK,QPSK\\
      Carrier frequency & $f_{c}$ & $2.437~\text{GHz}$\\
      USRP Bandwidth & $B$ & $ 20~\text{MHz}$\\
      I/Q sampling rate & $f_{s}$ & $20~\text{MSps}$\\
      \bottomrule
    \end{tabular}
    \label{tab:hw-parameters}
    \end{table}

\subsection{Transmission latency measurements}

\begin{figure}
    \centering
    \includegraphics[width=\linewidth]{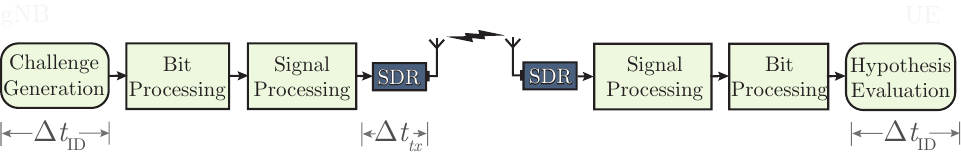}
    \vspace{-1em}
    \caption{Overview of latency measurements.}
    \label{fig:latency_meas_setup_id}
    \end{figure}

    In order to transmit the frames over the wireless channel using the SDR, once the bit and signal processing steps are carried out, the I/Q samples to be transmitted are passed onto the USRP hardware driver (UHD) via a shared buffer. We measure the time it takes for the UHD to empty this buffer using the parameters given in \cref{tab:hw-parameters} in order to calculate the transmission latency at the transmitter, as depicted in \cref{fig:latency_meas_setup_id}.
    
\subsection{Encoding and decoding latency measurements}

    We integrate the identification codes into our experimental setup, thus also transmitting the challenge.
    However, due to limitations between the software interfaces, such integration presents an overhead of a few seconds, which will need to be improved in future work. Therefore here we compare the computation time of the challenge only, estimated offline.
    
    The RS2 $0$-ID codes were implemented in Python and Sagemath, which uses three different C libraries (Givaro, NTL, PARI) depending on the finite field size. 
    For sizes less that 16 bits, Sagemath uses Givaro's Zech's table of logarithms to speed up  multiplications to times similar to  additions.

\section{Measurement results involving a single challenge}
\label{sec:single}

    \begin{figure}
        \centering
        \includegraphics[width=\columnwidth]{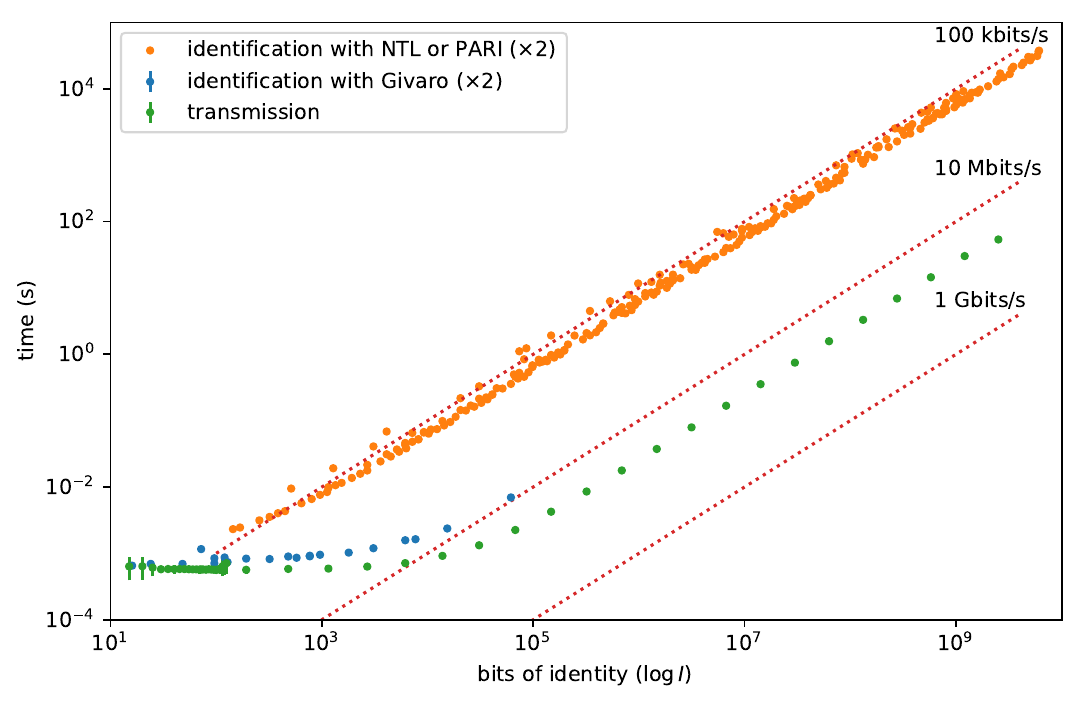}
        \vspace{-2.5em}
        \caption{Time needed to compute a challenge or to send the full identity.
        Since a tag must be computed both at the sender and at the receiver the computation time is doubled.
        The computation time is compared to the time needed to send the unique identifier of an identity (green data points).
        Some constant data rates (red) are plotted for reference.}
        \label{fig:timings}
        {\includegraphics[width=\columnwidth,trim=0 25 30 35, clip]{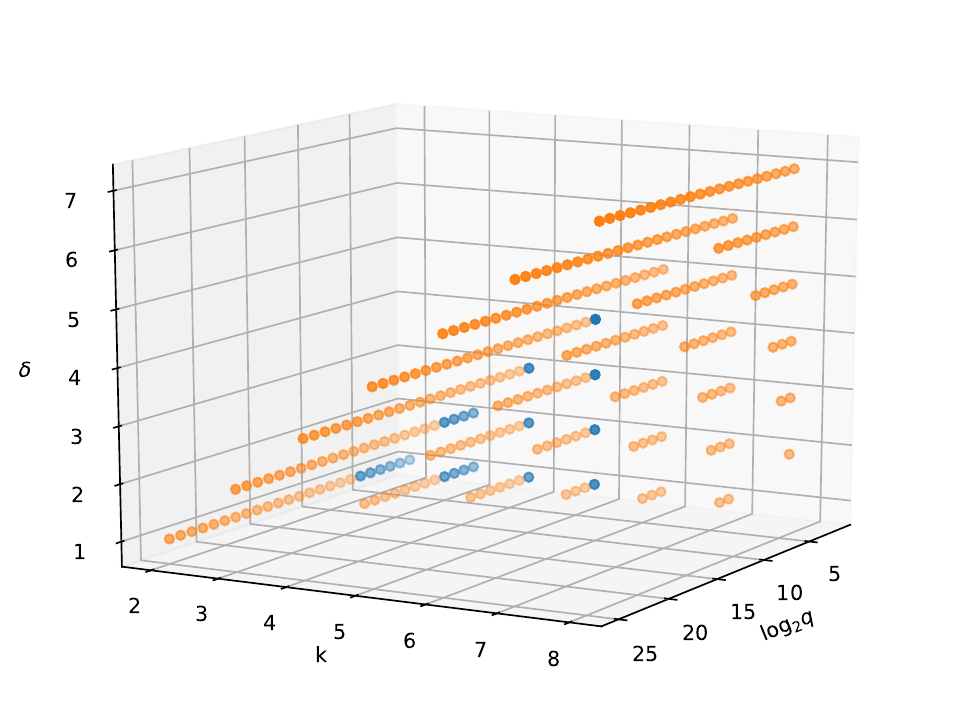}}
        \caption{The parameters explored in the data collection showing that the fast implementation is limited to low field sizes. }
    \end{figure}

\subsection{Trade-Offs in Identification vs. Transmission}

    In this section, we discuss and compare the case of generating a single instance of identification, i.e., sending the challenge $(r,t_i(r))$, against the scenario of sending the whole corresponding identity $i$ instead, as we would do in the classical transmission case.
    
    We consider a $[2^m,k,\delta]_\RST$ $0$-ID code which generates a challenge directly composed of $m \cdot (k+2)$ bits.
    The length of this $0$-ID code is the size of the first Reed-Solomon code, thus 
    $\log q^{k \cdot q^{k-\delta}} = m \cdot k \cdot  2^{m(k-\delta)}$ bits are needed to define an identity. 
    Due to time constraints, we limit ourselves to instances with a challenge computation time $2 \cdot  \Delta t_\mathrm{ID}$ of $2\cdot 10^4~\mathrm{s} \approx 6~\mathrm{h}$ (cf. \cref{fig:timings}). 
    The maximum number of identities achieved with this constraint is $[2^{13},7,5]_\RST$ at roughly $\sim5.7~\text{Gibits}$, where $1~\mathrm{Gibit}=2^{30}~\mathrm{bits}$, for which the challenge is just $117$ bits.
    
    In this respect, the identification code puts an undetectable load on the network.
    As displayed in \cref{fig:timings} and showcased for a few cases in \cref{tab:experiment_list}, this come at the cost of an exponential computational load at the sender and receiver.
    The parameter $\delta$ determines a trade-off between the computation time and the size of the identification code, for example $[2^{13},7,6]_\RST$ reduces exponentially the computation time but just decreases quadratically the number of identities. 
    Similarly for $m$, with the addition that $m$ is also a trade-off between the challenge length and the false accept probability (which scales as $\sim k/2^m + 1/2^{m\delta}$~\cite[Eq.~45]{DDF20}).

    \begin{table}[htbp]
    \caption{Sample subset of used parameters for a $[2^m,k,\delta]_\RST$ $0$-ID code}
    \centering
    \begin{tabular}{@{}ccccccc@{}}
    \toprule
    $m$ & $k$ & $\delta$ 
        & \begin{tabular}{@{}c@{}} 0-ID code \\ length \\ $m \cdot k \cdot  2^{m(k-\delta)}$ \end{tabular} 
        & \begin{tabular}{@{}c@{}} Challenge \\length \\ $m (k+2)$ \end{tabular} 
        & $2 \cdot \Delta t_\mathrm{ID}$ 
        & $\Delta t_\mathrm{tx}$\\
    \midrule
     3 & 3 & 2 & $72~\mathrm{bit}$ & $15~\mathrm{bit}$ &  $\sim 0.6~\mathrm{ms}$& $\sim 0.62~ \mathrm{ms}$\\
    25 & 3 & 2 & $2400~\mathrm{Mibit}$ & $125~\mathrm{bit}$ &  $\sim 4~\mathrm{h}$ & $\sim 53.7~ \mathrm{s}$\\
    13 & 7 & 6 & $ 728~\mathrm{Kibit}$ & $117~\mathrm{bit}$ &  $\sim 4~\mathrm{s}$ & $\sim 0.63~\mathrm{ms}$\\
    13 & 7 & 5 & $5824~\mathrm{Mibit}$ & $117~\mathrm{bit}$ &  $\sim 6~\mathrm{h}$ & --\\
    \bottomrule
    \end{tabular}
    \label{tab:experiment_list}
    \end{table}
    
\subsection{Latency Overhead}

    We compare the time spent generating and verifying the challenge $2 \cdot \Delta t_\mathrm{ID}$ (which for simplicity we calculate as twice the computation time) with the transmission time $\Delta t_\mathrm{tx}$ needed to send the $m \cdot k \cdot 2^{m(k-\delta)}$ bits of identity directly.

    As shown in \cref{fig:timings}, we find that for $m \cdot k< 16$, highlighted by the blue data points, the identification codes perform much faster than the other parameters, but most importantly its computation time is comparable, although still slower, to the latency introduced by transmission.
    Depending on the application, such small latency overhead might be justifiable in order to achieve the reduction in the network load obtained by sending these exponentially smaller challenges.
    For the largest of these identification codes, $[2^8, 5, 1]_\RST$, we have identities of size $60~\mathrm{Kibits}$ with challenges of size $56~\mathrm{bits}$.

\section{Conclusion} \label{sec:conclusion}
    Identification promises potential exponential advantages for applications where classical transmission is not strictly necessary.
    This is reminiscent of recommendation systems, where an exponential speedup is possible if one requires only samples rather than the whole recommendation vector~\cite{Tang19}.
    However, in order to showcase the advantages of identification, the computation time introduced by the $0$-ID codes needs to be accounted.
    
    In our experiments we have seen that identification can incur exponentially large computation times compared to transmission, which can affect the viability depending on the application.
    In particular, we find the importance of fast implementations of the involved operations, which, when available, allow for time performances comparable to transmission and thus allow to achieve the exponential reduction in the permit traffic with small compromise.
    
    The use of Zech's table allows the RS2 $0$-ID code to achieve computation times comparable with transmission times, but are not scalable to larger fields sizes.
    Instead, switching to $0$-ID codes based on Reed-Muller codes, namely adding variables to the polynomials, can increase the size of the identification code while keeping low the size of the finite field, and thus the computation time.
    This goal was indeed achieved during the revision of this manuscript in~\cite{SFD21}.
    Regardless, the small tags and the randomness being independent of the identities allow for other advantages like the use of common randomness and information-theoretic security. 

    Recently, some efficient functions have been shown to provide semantic security~\cite{WB21mosaics}, which is necessary to provide secure identification.
    Thus, future work will be devoted to exploring these security functions and more efficient $0$-ID codes.

\section*{Acknowledgements}
    We thank Sven Puchinger for his helpful and insightful comments.
    H. Boche is supported by the German Research Foundation (DFG) within the Gottfried Wilhelm Leibniz Prize under Grant BO 1734/20-1, and within Germany’s Excellence Strategy EXC-2111—390814868 and EXC-2092 CASA - 390781972.
    We also acknowledge support from the German Federal Ministry of Education and Research (BMBF)
    to H. Boche, L. Torres-Figueroa, W. Labidi, U. M\"onich, and A. Vlad-Costin under Grant 16KIS1003K and to C. Deppe and R. Ferrara under Grant 16KIS1005.

\end{document}